\documentclass[10pt,oneside,a4paper,reqno]{amsart}  
\usepackage{mathrsfs}

\usepackage{amssymb,
enumerate}
\usepackage{bbm}
\usepackage{geometry}                
\usepackage[T1]{fontenc}
\usepackage[english]{babel}
\usepackage{color}
\usepackage[colorlinks=true, linkcolor=magenta, citecolor=cyan, urlcolor=blue]{hyperref}

\renewcommand{\phi}{\varphi}

\newcommand{\meanv}[1]{\left\langle#1\right\rangle}

\def\be{\begin{equation}}
\def\ee{\end{equation}}
\def\bea{\begin{eqnarray}}
\def\eea{\end{eqnarray}}
\def\ni{\noindent}
\def\nn{\nonumber}

\DeclareMathOperator{\dist}{dist}

\newcommand{\E}[1]{\mathbb{E}\left[#1\right]}

\def\ie{\textit{i.e. }}

\def\a{\alpha}

\def\d{\delta}
\def\g{\gamma}

\def\b{\beta}

\def\s{\sigma}
\def\t{\tau}

\def\E{\mathbb{E}}

\def\N{\mathbb{N}}

\theoremstyle{plain}
\newtheorem*{theorem}{Theorem}

\numberwithin{equation}{section}

\definecolor{light}{gray}{.9}

\author{Giuseppe Genovese}
\address{Giuseppe Genovese: Institut f\"ur Mathematik, Universit\"at Z\"urich,
CH-8057 Z\"urich, Switzerland.}
\email{giuseppe.genovese@math.uzh.ch}

\author{Daniele Tantari}
\address{Daniele Tantari: Centro Ennio de Giorgi, Scuola Normale Superiore, Piazza dei Cavalieri 3, I-56100 Pisa (Italy).}
\email{daniele.tantari@sns.it}

\title[multipartite random energy models]
{Overlap synchronisation in multipartite random energy models}
\date{\today}

\begin{document}
\begin{abstract}
In a multipartite random energy model, made of a number of coupled GREMs, we determine the joint law of the overlaps in terms of the ones of the single GREMs. This provides the simplest example of the so-called {\em overlap synchronisation}.

\vspace{0.5cm}
\ni {\bf MSC:} 82B44, 60G55, 60K35.
\end{abstract}

\pagestyle{plain}

\maketitle

\section{Introduction} 

The overlap synchronisation phenomenon was recently introduced by Panchenko in \cite{pan} for multipartite spin glasses \cite{BCMT}. 
The study of such systems is of primary interest, because of the applications in neural network theory and statistical inference \cite{mez,lettera}: {\em e.g.} the Hopfield model, the restricted Boltzmann machine and the perceptron are examples of bipartite spin glasses. The lack of  convexity prevents to apply directly to multipartite spin-glasses some useful techniques developed for the Sherrington-Kirpatrick model (\ie interpolation bounds \cite{guerra}), calling for new ideas. 

In this note we investigate a multipartite random energy model, originally studied for the bipartite case in \cite{franz}, obtained coupling each level of $M$ distinct generalised random energy models (GREMs). We show the joint law of the overlaps to have a direct expression in terms of the ones of the single GREMs. This provides a simple example of overlap synchronisation.  

The model is defined as follows. Let $N,M\in\N$, $\kappa\in\{1,\dots,M\}$ and $N_\kappa \in\N$ with $\sum_\kappa N_\kappa=N$, $\a^{(\kappa)}:=N_\kappa/N$.
For each configuration $\s\in\Sigma_N:=\{1,\dots,2^N\}$ we can identify 
$\s=(\mu_{(1)},\dots,\mu_{(M)})$, $\mu_{(\kappa)}\in\{1,\ldots,2^{N_\kappa}\}$.
We divide each part respectively into $K_1,\dots K_M$ hierarchical levels. For each level $j$ of the hierarchy, each group of configurations is divided in $2^{N_{\kappa,j}}$ further subgroups indexed by $\mu_{(\kappa,j)}$,
with of course $\sum_j N_{\kappa,j}=N_\kappa$ and $\varsigma_{\kappa,j}:=N_{\kappa,j}/N_\kappa$, $j\in\{1,\dots,K_\kappa\}$. Each configuration can be thought of as a $M$-ple $\s=(\mu_{(1)},\dots,\mu_{(M)})$ or as a $\left(\prod_{\kappa}K_\kappa\right)$-ple $\s=(\mu_{(1,1)}\dots\mu_{(1,K_1)},\dots,\mu_{(M,1)},\dots\mu_{(M,K_M)})$. This multipartite setting brings a somewhat heavy notation. To lighten it a little we let
$$
\ell_{\kappa,j}:=\mu_{(\kappa,1)}\dots\mu_{(\kappa,j)}\,\quad \mbox{(note $\ell_{\kappa,1}=\mu_{(\kappa,1)}$ and $\ell_{\kappa,K_\kappa}=\mu_{(\kappa)}$)}
$$ 
label the configurations in the $j$-th level of the $\kappa$-th tree. With a slight abuse of notation we will denote with the same symbol also the set of such configurations (the correct meaning will be always clear from the context). 
We attach to each couple of levels Gaussian centred r.vs $J^{(\kappa,j)}_{\ell_{\kappa,j}}$, 
and $J^{(\kappa_1,j_1)(\kappa_2,j_2)}_{\ell_{\kappa_1,j_1}\ell_{\kappa_2,j_2}}$ with 
\bea
E\left[J^{(\kappa,j)}_{\ell_{\kappa,j}}J^{(\kappa,j)}_{\ell'_{\kappa,j}}\right]&=& \d_{\ell_{\kappa,j},\ell'_{\kappa,j}}\,,\nn\\
E\left[J^{(\kappa_1,j_1)(\kappa_2,j_2)}_{\ell_{\kappa_1,j_1}\ell_{\kappa_2,j_2}}J^{(\kappa_1,j_1)(\kappa_2,j_2)}_{\ell'_{\kappa_1,j_1}\ell'_{\kappa_2,j_2}}\right]&=&\d_{\ell_{\kappa_1,j_1},\ell'_{\kappa_1,j_1}}\d_{\ell_{\kappa_2,j_2},\ell'_{\kappa_2,j_2}}\,.\nn
\eea
The levels interact via the following Hamiltonian 
\be\label{eq:HGREM}
H_N(\s):=-\sqrt{\frac{N}{2}}\left[\sum_{\kappa=1}^M\a^{(\kappa)}\sum_{j=1}^{K_\kappa}a^{(\kappa)}_j J^{(\kappa,j)}_{\ell_{\kappa,j}}+\sqrt{2\a^{(1)}\ldots\a^{(M)}}\sum_{(\kappa_1,\kappa_2)}\sum_{j_1=1}^{K_{\kappa_1}} \sum_{j_2=1}^{K_{\kappa_2}}c^{(\kappa_1,\kappa_2)}_{j_1,j_2}J^{(\kappa_1,j_1)(\kappa_2,j_2)}_{\ell_{\kappa_1,j_1}\ell_{\kappa_2,j_2}}\right]\,,
\ee
with 
$$
\sum_j^{K_\kappa}a^{(\kappa)}_j=\sum_{j_1=1}^{K_{\kappa_1}}\sum_{j_2=1}^{K_{\kappa_2}}c^{(\kappa_1,\kappa_2)}_{j_1,j_2}=1\,,\quad \forall\,\, \kappa,\kappa_1,\kappa_2\in\{1,\ldots,M\}\,.
$$
We can introduce $M$ different partial overlaps between two different configurations $\s\neq\s'$ as
\be
\t_\kappa=\tau_\kappa(\s,\s'):=\inf\left\{j\geq0\,:\,\mu_{(\kappa,j+1)}\neq{\mu'}_{(\kappa,j+1)}\right\}\,,\quad \kappa=1\ldots,M\,,
\ee
and $\t_\kappa=K_\kappa$ if $\s=\s'$. Then a direct computation gives
$$
E[H_N(\s)H_N(\s')]=\frac N2 \sum_{\kappa=1}^M{\a^{(\kappa)}}^2\sum_{j=1}^{\t_\kappa}{a^{(\kappa)}_j}^2+2\a^{(1)}\ldots\a^{(M)}\sum_{j_1,j_2=1}^{\min\left(\t_{\kappa_1},\tau_{\kappa_2}\right)}{c^{(\kappa_1,\kappa_2)}_{j_1,j_2}}^2\,. 
$$
It is somehow convenient to set the overlaps in $[0,1]$: we introduce $M$ sequences of numbers in $[0,1]$
$$
0=q^{(\kappa)}_0<q^{(\kappa)}_1<\cdots<q^{(\kappa)}_{K_\kappa}=1\,,\qquad \kappa\in\{1,\dots,M\}\,
$$
and put $q_\kappa=q_{\kappa}(\s,\s'):=q^{(\kappa)}_{\tau_{\kappa}}$. We also define the total overlap to be $q_{tot}:=\sum_{\kappa=1}^M\a_\kappa q_\kappa$.
As customary for $\b>0$ ($-\frac1\b$) the free energy is given by
\be\label{eq:A}
A_N(\b):=\frac1N \log \sum_{\s} e^{-\b H_N(\s)}\,,\qquad A(\b):=\lim_NA_N(\b)\,. 
\ee
Of course as a consequence of Talagrand inequality $A_N(\b)$ is self-averaging as $N\to\infty$, so we can always take the expectation w.r.t. the disorder, when needed. Here and further we denote by $P_{N,\b}$ the Gibbs distribution associated to the model and by $\meanv{\cdot}_{N,\b}$ the quenched average of observables (we drop the subscript $N$ in the thermodynamic limit) and
$
x_\kappa(q):=P_{\b}\left(q_{\kappa}\leq q\right)
$, $x_{tot}(q):=P_{\b}\left(q_{tot}\leq q\right)$.

Our main result is
\begin{theorem}\label{TH}
Let $\upsilon$ be a random variable uniformly distributed in $[0,1]$. Then
\be\label{eq:TH}
\left(q_1,\ldots,q_M\right)\overset {d}{=} (x^{-1}_1\left(\upsilon),\ldots,x^{-1}_M(\upsilon)\right)\,.
\ee
\end{theorem}

This result can be given also in terms of the total overlap (as in \cite{pan}):
\be\label{eq:TH2}
\left(q_1,\ldots,q_M\right)\overset {d}{=} \left(x^{-1}_1	\circ x_{tot}(q_{tot}),\ldots,x^{-1}_M\circ x_{tot}(q_{tot})\right)\,.
\ee

A larger class of non-hierarchical random energy models including the one under consideration was studied by Bolthausen and Kistler in \cite{BK1,BK2}. We shall make use of some crucial ideas from those two papers, in which the so-called {\em Parisi picture} is proved. A more precise formulation of the results in \cite{BK1,BK2} will be given below.

\vspace{0.1cm}

\section{More on the Model} 
Prior to give the proof, it is convenient to discuss a little more the model. What follows is in a good part heuristics and rigorous proofs can be found in \cite{BK1, BK2}. 

For $M=1$ and $K_1=1$, we simply recover the usual REM. We shortly summarise some basic features of this well-known model. The model has a phase transition at $\b_c:=2\sqrt{\log2}$, so that $x(q)=1$ for $\b\leq\b_c$ and $x(q)=\b_c/\b$ otherwise. The free energy reads
$$
A_{REM}(\b):=1_{\{\b\leq \b_c\}}\log2\left(1+\frac{\b^2}{\b_c^2}\right)+1_{\{\b>\b_c\}}2\log2\frac{\b}{\b_c}\,. 
$$
Next consider for simplicity the bipartite model ($M=2$) with $K_1=K_2=1$, defined by the Hamiltonian (we set $a_1^{(1)}=a$, $a_1^{(2)}=b$ and $\a^{(1)}=\a$) 
\be\label{eq:HREM}
H_N(\s):=-\sqrt{\frac{N}{2}}\left[\a aJ^{(1,1)}_{\mu_{(1)}}+(1-\a)bJ^{(2,1)}_{\mu_{(2)}}+\sqrt{2\a(1-\a)}cJ^{(1,1)(2,1)}_{\mu_{(1)}\mu_{(2)}}\right]\,
\ee
(this was analysed also in \cite{AK} by a slightly different perspective). 
If we assume for definiteness $\a a^2>(1-\a)b^2$, there are two possibilities: either $\a a^2\leq (1-\a)b^2+2\a c^2$ or $\a a^2 > (1-\a)b^2+2\a c^2$. The first case is less interesting and we focus on the second one. 
At very high temperature everything is ergodic and the free energy coincides with the annealed one. As $\b>\b_1:=2\sqrt{\log2}/a\sqrt{\a}$, the $\mu_{(1)}$-subset {\em freezes}, \ie its relative entropy goes to zero (as in the first transition in a GREM \cite{DG}) and one can show that 
$$
P^{(1)}_{N,\b}(\mu_{(1)};\b):=Z^{-1}\sum_{\mu_{(2)}} e^{-\b H_N(\s)}\rightharpoonup PD(0,\b_1/\b)\,,
$$
where $Z$ is a normalisation factor and $PD(0,x)$ denotes the law of a normalised Poisson point process with intensity $\rho(t)=xt^{-x-1}$ or Poisson-Dirichlet distribution. In this regime for any $q,p>0$ $x_1(q)=\b_1/\b$, while $P_\b(q_{2}\geq p)=P_\b(q_{1}\geq q,q_{2}\geq p)=0$. The free energy is a convex combination (with $\a$) of two REMs, one on the $\mu_{(1)}$ subset at low temperature and the other on the rest of the system at high temperature. As $\b$ increases further, the total entropy vanishes for $\b>\b_2:=2\sqrt{\log2}/\sqrt{(1-\a)b^2+2\a c^2}$ and the whole Gibbs measure converges toward a Poisson-Dirichlet process
$$
P_{N,\b}(\s;\b)\rightharpoonup PD(0,\b_2/\b)\,,
$$
The free energy is the convex combination of two REMs at low temperature.
$x_1(q)$ is unchanged, but $1-x_2(p)=P_\b(q_{1}\geq q,q_2\geq p)=1-\b_2/\b$. Note 
$$
P_\b\left(q_{1}\geq q,q_2\geq p\right)=\min\left(P_\b(q_{1}\geq q)\,,\,\,P_\b(q_{2}\geq p)\right)
$$ 
for any $\b$. We remark that, since the overlaps take value in $\{0,1\}$ in this simple case, $P_\b(q_{1}\geq q)=P_\b(q_{1}= 1)$ and $P_\b(q_{2}\geq p)=P_\b(q_{2}= 1)$ (as $q,p>0$). Therefore the first system starts freezing at higher temperature: if the second system is frozen, then also the first one is so (as in a two-level GREM \cite{DG}). The whole picture is summarised as follows

\

\begin{center}
\begin{tabular}{c|c|c|c}
$\b$&$A(\b)$&$\big\langle q_{1}\big\rangle_\b$&$\big\langle q_{2}\big\rangle_\b$\\ \hline
$[0,\b_1]$&$\log2\left(1+\a\frac{\b^2}{\b_1^2}+(1-\a)\frac{\b^2}{\b_2^2}\right)$&
$0$&$0$\\ 
\hline $(\b_1,\b_2)$&$\log2\left(2\a\frac{\b}{\b_1}+(1-\a)\left(1+\frac{\b^2}{\b_2^2}\right)\right)$ &
$1-\frac{\b_1}{\b}$&$0$\\
\hline $[\b_2,\infty)$&$2\log2\left(\a\frac{\b}{\b_1}+(1-\a)\frac{\b}{\b_2}\right)$ &
$1-\frac{\b_1}{\b}$&$1-\frac{\b_2}{\b}$\\ \hline
\end{tabular}
\end{center}

\

Therefore, albeit not inbuilt in the model, a GREM-like hierarchical structure naturally emerges. A way to visualise that in the general model defined by the Hamiltonian (\ref{eq:HGREM}) is as follows. Recall that the $\ell_{\kappa,j}$, $\kappa\in\{1,\ldots,M\}$, $j\in\{0,\dots,K_\kappa\}$, denote the configurations up to the $j$-th level of the $\kappa$-th GREM. Then the phase space is naturally coarse-grained by the class of sets $\{\ell_{\kappa_1,j_1},\ell_{\kappa_2,j_2}\}^{j_1=1,\dots, K_{\kappa_1}}_{j_2=1,\dots, K_{\kappa_2}}$. We think of each level now as an atom and we can consider the power set  
$$
\wp:=\wp\{\ell_{1,1},\ldots,\ell_{1,K_1},\ldots,\ell_{M,1},\ldots,\ell_{M,\kappa_M}\}\,.
$$
According to \cite{BK1,BK2} a {\em chain} $\Gamma$ is defined to be an increasing (finite) sequence of $K\leq\sum_{\kappa} K_\kappa$ sets in $\wp$ $\Gamma=\{\Gamma_n\}_{n=0,\dots,K}$ so that
$$
\Gamma_n\in\wp\,,\,\, \Gamma_n\subset\Gamma_{n+1}\,, \,\,\Gamma_0=\emptyset\,,\,\,\Gamma_K=\{\ell_{1,0},\ldots,\ell_{1,K_1},\ldots,\ell_{M,1},\ldots,\ell_{M,\kappa_M}\}\,.$$
To each $\Gamma$ we associate two sequences $\{\a_n\}_{n=1,\dots,K}$ and $\g:=\{\g_n\}_{n=1,\dots,K}$. The $\a_n$ represent the relative sizes of the $\Gamma_n$ 
$$\a_n:=\frac{\log_2\left|\bigcup_{i,j:\ell^i_j\in \Gamma_n\setminus \Gamma_{n-1}} \ell^i_j\right|}{N}\,,
$$ easily computed from the numbers $\a^{(\kappa)}$ and $\varsigma_{\kappa,j}$; the $\g_n$ are variances defined by
\be\label{eq:gamma-n}
\gamma^2_n:=\sum_{\kappa=1}^M{\a^{(\kappa)}}^2\sum_{j\,:\,\ell_{\kappa,j}\in\Gamma_{n}\setminus\Gamma_{n-1}} (a^{(\kappa)}_{j})^2+\sum_{\overset{(\kappa_1,\kappa_2)\,,\,(j_1,j_2)\,:}{\,:\,\{\ell_{\kappa_1,j_1},\ell_{\kappa_2,j_2}\}\in\Gamma_{n}\setminus\Gamma_{n-1}}} 2\a_{\kappa_1}\a_{\kappa_2}(c^{(\kappa_1,\kappa_2)}_{j_1,j_2})^2\,.
\ee
From $\a_n$ and $\g_n$ we can build another sequence of critical inverse temperatures $\{\b_n\}_{n=1,\dots,K}$, $\b_n:=\sqrt{\a_n\log 2}\gamma_n^{-1}$. In general $\{\b_n\}_{n=1,\dots,K}$ is not monotone, but we can conveniently confine our attention to those chains for which $\b_1\leq\b_2\leq\ldots\leq\b_K$. We denote by ${\mathcal T}$ the set of such chains.

To fix the ideas, let us consider again a bipartite REM with $K_1,K_2$ levels. The Hamiltonian reads as
\be\label{eq:Hbip}
H_N(\s)=
-\sqrt{\frac{N}{2}}\left[\a\sum_{j=1}^{K_1}a^{(1)}_j J^{(1,j)}_{\ell_{1,j}}+(1-\a)\sum_{j=1}^{K_2}a^{(2)}_j J^{(2,j)}_{\ell_{2,j}}+\sqrt{2\a(1-\a)}\sum_{j_1=1}^{K_{1}} \sum_{j_2=1}^{K_{2}}c_{j_1,j_2}J^{(1,j_1)(2,j_2)}_{\ell_{1,j_1}\ell_{2,j_2}}\right]\,
\ee

For a given $\Gamma \in\mathcal T$ of length $K$, we set for $n=1,\ldots,K$
\bea
H_n&:=&-\sqrt{\frac{N}{2}}\left[\a\sum_{j\,:\, \ell_{1,j}\in\Gamma_n/\Gamma_{n-1}}a^{(1)}_j J^{(1,j)}_{\ell_{1,j}}+(1-\a)\sum_{j\,:\, \ell_{2,j}\in\Gamma_n/\Gamma_{n-1}}a^{(2)}_j J^{(2,j)}_{\ell_{2,j}}\right.\nn\\
&+&\left.\sqrt{2\a(1-\a)}\sum_{(j_1,j_2)\,:\,\ell_{1,j_1},\ell_{2,j_2}\in\Gamma_{n}\setminus\Gamma_{n-1}}c_{j_1,j_2}J^{(1,j_1)(2,j_2)}_{\ell_{1,j_1}\ell_{2,j_2}}\right]\,\,,\nn
\eea
so that we can decompose the Hamiltonian (\ref{eq:Hbip}) according to
\be\label{eq:decH}
H_N(\s)=\sum_{n=1}^{K} H_n\,,
\ee
and the partition function can be written as
$$
Z_N(\b)=\sum_{\{\Gamma_1\}}e^{-\b H_1}\sum_{\{\Gamma_2/\Gamma_1\}}e^{-\b H_2}\dots \sum_{\{\Gamma_n/\Gamma_{n-1}\}}e^{-\b H_n}\,.
$$
Now we see the following scenario. At $\b$ small enough the annealed approximation holds and the overlaps are set to zero. Then $\b$ increases, $\b>\b_1$, and the configurations in $\Gamma_1$ {\em freeze}. In fact $H_2$ depends on configurations in $\Gamma_2/\Gamma_1$, \ie $H_1$ and all the other addenda in the r.h.s. of (\ref{eq:decH}) become independent as $N\to\infty$. Thus the partition function asymptotically factorises
$$
Z_N(\b)\simeq\sum_{\{\Gamma_1\}}e^{-\b H_1}\sum_{\{\Sigma/\Gamma_1\}}e^{-\b (H-H_1)}
$$
as two independent REMs: the first one on the space of configurations $\Gamma_1$ is at low temperature, the second one on the remaining configuration space is at high temperature (with the right variance $\sqrt{\sum_{n\geq2}\gamma^2_n}$). The free energy is a convex combination w.r.t. $\a_1$ (\ie the relative size of $\Gamma_1$)  of these two REMs. 
As in the previous example, we have convergence of the marginalised Gibbs measure to a Poisson-Dirichlet distribution
$$
P^{(1)}_{N,\b}(\Gamma_1;\b):=Z_1^{-1}\sum_{\Sigma/\Gamma_1} e^{-\b H(\s)}\rightharpoonup PD(0,\b_1/\b)\,,
$$
with $Z_1$ an opportune normalisation. Since $H_1$ and $H_2$ remain independent for all $\b>\b_1$ we can iterate this procedure: for instance as $\b>\b_2$ also $\Gamma_2$ {\em freezes} and $H_2$ becomes asymptotically independent on $H_3$; thus the partition function is factorised as
$$
Z_N(\b)\simeq\sum_{\{\Gamma_1\}}e^{-\b H_1}\sum_{\{\Gamma_2/\Gamma_1\}}e^{-\b H_2}\sum_{\{\Sigma/\Gamma_2\}}e^{-\b (H-H_1-H_2)}\,.
$$
These are three independent REMs on configurations $\Gamma_1$, $\Gamma_2/\Gamma_1$ and $\Sigma/\Gamma_2$, the associated free energy is given by a convex combination of the low-temperature free energies of the first two REMs and the high-temperature free energy of the third one and
$$
P^{(2)}_{N,\b}(\Gamma_2;\b):=Z_2^{-1}\sum_{\Sigma/\Gamma_2} e^{-\b H(\s)}\rightharpoonup PD(0,\b_2/\b)\,,
$$
$Z_2$ is a normalisation factor. Going on this way we recover the free energy and the Gibbs measure as a GREM-like structure along the chain. At zero temperature the free energy of the model is just the convex combination of those of REMs at low temperature, each defined on an element of the chain. This construction can be made for every chain in $\mathcal T$. Of course for fixed $\b$, the more REMs are at low temperature, the higher is the free energy. According to this criterion one can select the chain along which the free energy is maximal. By the above construction it should be clear that such a chain, here denoted by $\Gamma^*$, is unique. 

The results of \cite{BK1,BK2} (for the case of our interest) can be precisely formulated as follows. Here $\Gamma\in\mathcal T$ and $A_{GREM}(\Gamma;\b)$ denotes the GREM pressure computed along $\Gamma$:
\be
A_{GREM}(\Gamma;\b):=\sum_{n=1}^K 1_{\{\b\leq \b_n\}}\a_n\log2\left(1+\frac{\b^2}{\b_n^2}\right)+1_{\{\b> \b_n\}}2\a_n\log2 \frac{\b}{\b_n}\,. 
\ee
\begin{theorem}[Bolthausen and Kistler]
The following holds:
\begin{enumerate}
\item[i)]
$
\lim_NA_N(\b)=\lim_N\E[A_N(\b)]=A(\b)=\min_{\g\in\mathcal T}\left(A_{GREM}(\g;\b)\right)\,.
$\\
\item[ii)] Ultrametricity: there is a $\b^*$ such that for each triad of configurations $(\s,\s',\s'')\in\Sigma^3$
$$
\lim_NP_{N,\b}\left(\dist(\s,\s')\leq\max\{\dist(\s,\s''),\dist(\s',\s'')\}\right)=1\,,\quad \forall \b>\b^*\,. 
$$
\end{enumerate}
\end{theorem}

For a thorough discussion of point $ii)$ we refer to the original work, but is worth mentioning it comes from Theorem 3 in \cite{BK2}, on which we roughly report: for $\b>\b_n$ $P_{N,\b}^{(n)}\rightharpoonup PD(0,\b_n/\b)$ and the overlap converges in distribution to the Bolthausen-Sznitman coalescence along the optimal chain $\Gamma^*$ (the $\b_n$'s being associated to $\Gamma^*$); moreover the overlap and the Gibbs measure are asymptotically independent. 

\vspace{0.2cm}

\section{Proof} 
Now we are ready to give the proof of our statement. For simplicity we keep working mostly in the bipartite case.
We convey to fix the optimal chain $\Gamma^*$ once for all. The sequences $\{\a_n\}$, $\{\gamma_n\}$ and $\{\b_n\}$ will be always referred to $\Gamma^*$. 

A direct computation from (\ref{eq:HGREM}) and (\ref{eq:A}) yields
\bea
P_{N,\b}\left(q_{1}\geq q^{(1)}_j\right)&=&\meanv{1_{\{\ell_{1,j}={\ell'_{1,j}}\}}}_{N,\b}=1-\frac{2}{a^{(1)}_j\b^2\a^2}\partial_{a^{(1)}_j} A_N\,,\nn\\
P_{N,\b}\left(q_{2}\geq q^{(2)}_j\right)&=&\meanv{1_{\{\ell_{2,j}={\ell'_{2,j}}\}}}_{N,\b}=1-\frac{2}{a^{(2)}_j\b^2(1-\a)^2}\partial_{a^{(2)}_j} A_N\,,\nn\\
P_{N,\b}\left(q_{1}\geq q^{(1)}_{j_1},\,q_{2}\geq q^{(2)}_{j_2}\right)&=&\meanv{1_{\left\{(\ell_{1,j_1},\,\ell_{2,j_2})=(\ell'_{1,j_1},\,\ell'_{2,j_2})\right\}}}_{N,\b}=1-\frac{1}{c_{j_1j_2}\b^2\a(1-\a)}\partial_{c_{j_1j_2}} A_N\,.\nn
\eea
On the other hand the limiting free energy is a convex combination of REM ones along $\Gamma^*$, thus its derivatives can be explicitly computed.  We set 
$$
n_1(j):=\min\{n\,:\,\ell_{1,j}\in\Gamma^*_n\}\,,\quad n_2(j):=\min\{n\,:\,\ell_{2,j}\in\Gamma^*_n\}\,,\quad n(j_1,j_2):=\max(n_1(j),n_2(k))\,.
$$
Then
\be
\partial_{a^{(1)}_j} A=
\begin{cases}
\b^2\frac{\a^2a^{(1)}_j}{2}&\b<\b_{n_1(j)}\\
\b\sqrt{\a_{n_1(j)}\log 2}\a^2\frac{a^{(1)}_j}{\gamma_{n_1(j)}}&\b\geq\b_{n_1(j)}\,,
\end{cases}
\ee
\be
\partial_{a^{(2)}_j} A=
\begin{cases}
\b^2\frac{(1-\a)^2a^{(2)}_j}{2}&\b<\b_{n_2(j)}\\
\b\sqrt{\a_{n_2(j)}\log 2}(1-\a)^2\frac{a^{(2)}_j}{\gamma_{n_2(j)}}&\b\geq\b_{n_2(j)}\,,
\end{cases}
\ee
\be
\partial_{c_{jk}} A=
\begin{cases}
\b^2\a(1-\a)c_{jk}&\b<\b_{n(j_1,j_2)}\\
\b\sqrt{\a_{n(j_1,j_2)}\log 2}2\a(1-\a)\frac{c_{jk}}{\gamma_{n(j_1,j_2)}}&\b\geq\b_{n(j_1,j_2)}\,.
\end{cases}
\ee
As $N\to\infty$ the two expressions for the derivatives must be equal (the exchange of the limit and the derivative can be justified for instance using Theorem 3 in \cite{BK2} and the concavity of the free energy). Hence if $n_1(j_1)=n_2(j_2)=\bar n$
\be
P_\b\left(q_{1}\geq q^{(1)}_j\right)=P_\b\left(q_{2}\geq q^{(2)}_j\right)
=P_\b\left(q_{1}\geq q^{(1)}_{j_1},\,q_{2}\geq q^{(2)}_{j_2}\right)=
\begin{cases}
0&\b<\b_{\bar n}\\
1-\frac{\b_{\bar n}}{\b}&\b\geq \b_{\bar n}\,.
\end{cases}\label{eq:P->n=n=n}
\ee
Otherwise we have
\be\label{eq:P->q}
P_\b\left(q_{1}\geq q^{(1)}_j\right)=\begin{cases}
0&\b<\b_{n_1(j)}\\
1-\frac{\b_{n_1(j)}}{\b}&\b\geq \b_{n_1(j)}\,,
\end{cases}
\ee
\be\label{eq:P->p}
P_\b\left(q_{2}\geq q^{(2)}_j\right)=\begin{cases}
0&\b<\b_{n_2(j)}\\
1-\frac{\b_{n_2(j)}}{\b}&\b\geq \b_{n_2(j)}\,
\end{cases}
\ee
and
\be\label{eq:P->pq}
P_\b\left(q_{1}\geq q^{(1)}_{j_1},\,q_{2}\geq q^{(2)}_{j_2}\right)=\begin{cases}
0&\b<\b_{n(j_1,j_2)}\\
1-\frac{\b_{n(j_1,j_2)}}{\b}&\b\geq \b_{n(j_1,j_2)}\,.
\end{cases}
\ee
As $1-\b_n/\b$ is decreasing in $n$, formulas (\ref{eq:P->n=n=n}) and (\ref{eq:P->q}), (\ref{eq:P->p}), (\ref{eq:P->pq}) establish directly
\be\label{eq:P=min}
P_\b\left(q_{1}\geq q^{(1)},\,q_{2}\geq q^{(2)}\right)=\min\left(P_\b(q_{1}\geq q^{(1)})\,,\,\,P_\b(q_{2}\geq q^{(2)})\right)\,. 
\ee
Let now $\upsilon\sim U(0,1)$. We have
\bea
P_\b\left(q_{1}\geq q^{(1)}, q_{2}\geq q^{(2)}\right)&=&\min\left[P_\b( q_{1}\geq q^{(1)}\,)\,,\, \, P_\b( q_{2}\geq q^{(2)}\,)\right]\nn\\
&=&P_\b\left(\upsilon\leq\min\left[P_\b( q_{1}\geq q^{(1)}\,)\,,\, \, P_\b( q_{2}\geq q^{(2)}\,)\right]\right)\nn\\
&=&P_\b\left(x_1^{-1}(1-\upsilon)\geq q^{(1)}, x_2^{-1}(1-\upsilon)\geq q^{(2)}\right)\nn\,,
\eea 
from which (\ref{eq:TH}) is readily deduced for $M=2$.

The generalisation to the multipartite case is immediate. We have for every $(\kappa_1,\kappa_2)\in\{1,\ldots,M\}^2$
\be\label{eq:segna}
P_\b\left(q_{\kappa_1}\geq q^{(\kappa_1)},\,q_{\kappa_2}\geq q^{(\kappa_2)}\right)=\min\left(P_\b(q_{\kappa_1}\geq q^{(\kappa_1)})\,,\,\,P_\b(q_{\kappa_2}\geq q^{(\kappa_2)})\right)\,,
\ee
whence, proceeding as above, $(q_{\kappa_1},q_{\kappa_2})\overset{d}{=}(x_1^{-1}(\upsilon), x_2^{-1}(\upsilon))$, $\upsilon\sim U(0,1)$. Therefore we have mutual synchronisation for every couple $(\kappa_1,\kappa_2)$, which is enough to obtain (\ref{eq:TH}) for any $M$. 
Moreover a simple computation from (\ref{eq:segna}) also gives
$$
P_\b\left(q_{\kappa}\geq q,\,q_{tot}\geq Q\right)=\min\left(P_\b(q_{\kappa}\geq q)\,,\,\,P_\b(q_{tot}\geq Q)\right)\,,
$$
for any $\kappa=1,\ldots,M$, thus for $\upsilon\sim U(0,1)$
$$
(q_1,\ldots,q_M,q_{tot})\overset {d}{=} (x_1^{-1}(\upsilon),\ldots,x_M^{-1}(\upsilon),x_{tot}^{-1}(\upsilon))\,,
$$
where we note $x^{-1}_{tot}(z)=\sum_{\kappa=1}^M\a_\kappa x_\kappa^{-1}(z)$. Then (\ref{eq:TH2}) easily follows.

\

{\bf Acknowledgements} This research was supported through the programme {\em Research in Pairs} by the Mathematisches Forschungsinstitut Oberwolfach in November 2016. G.G. is supported by the NCCR SwissMAP, D.T. is supported by GNFM-Indam.
We thank E. Bolthausen for some valuable discussions and S. Franz for a useful correspondence on the paper \cite{franz}.


\end{document}